\documentclass[preprint,amssymb,amsmath,aps,prb,showpacs]{revtex4}
\usepackage{graphicx}

\begin{document}

\title{Mesoscopic BCS pairing in the repulsive 1d-Hubbard model}

\author{Luigi Amico}
\affiliation{MATIS, INFM CNR}  
\affiliation{Dipartimento di Metodologie Fisiche e Chimiche (DMFCI),
Universit\`a di Catania, viale A. Doria 6, I-95125 Catania, Italy}
\email{lamico@dmfci.unict.it}
\homepage{http://fisica.ing.unict.it/~lamico}

\author{Andrea Mastellone}
\affiliation{MATIS, INFM CNR}  
\affiliation{Dipartimento di Metodologie Fisiche e Chimiche (DMFCI),
Universit\`a di Catania, viale A. Doria 6, I-95125 Catania, Italy}
\email{andrea@femto.dmfci.unict.it}
\homepage{http://fisica.ing.unict.it/~andrea}

\author{Andreas Osterloh}
\affiliation{Institut f\"ur Theoretische Physik, Universit\"at Hannover,
30167 Hannover, Germany}
\email{andreaso@itp.uni-hannover.de}
\homepage{http://www.itp.uni-hannover.de/~andreaso/}

\date{\today}

\begin{abstract}
We study mesoscopic pairing in the one dimensional repulsive 
Hubbard model and its interplay with the BCS model in the canonical ensemble. 
The key tool is comparing the 
Bethe ansatz equations of the two models in the limit of small Coulomb 
repulsion. For the ordinary 
Hubbard interaction the BCS Bethe equations with infinite  pairing 
coupling are recovered; a finite pairing 
is obtained by considering a further density-dependent  
phase-correlation in the hopping amplitude of the Hubbard model. 
We find that spin degrees of freedom 
in the Hubbard ground state are arranged in a state 
of the BCS type, where  the Cooper-pairs form an non-condensed liquid on 
a ``lattice'' of single particle energies provided by the Hubbard charge 
degrees of freedom; the condensation in the BCS ground state corresponds 
to Hubbard excitations constituted by a sea of spin singlets.  
\end{abstract}


\maketitle


\def\bea{\begin{eqnarray}}
\def\eea{\end{eqnarray}}
\def\beq{\begin{equation}}
\def\eeq{\end{equation}}
\def\up{\uparrow}
\def\down{\downarrow}
\def\ket#1{ | #1 \rangle}


\def\al{\alpha}
\def\be{\beta}
\def\ga{\gamma}
\def\Ga{\Gamma}
\def\de{\delta}
\def\De{\Delta}
\def\ep{\epsilon}
\def\vep{\varepsilon}
\def\la{\lambda}
\def\La{\Lambda}
\def\sig{\sigma}
\def\Sig{\Sigma}
\def\vphi{\varphi}
\def\om{\omega}
\def\Om{\Omega}


\section{Introduction}

Intensive analysis has been devoted to trace pairing phenomena in strongly 
interacting electronic models\cite{NOZIER,YANG, CAPONE}. Most of the 
motivations come from studies in high-$T_c$ superconductivity for which it is 
highly desirable that pairing mechanisms are automatically encoded 
in the interaction. The Hubbard model constitutes the prototype of the 
class of models of interest in the subject\cite{HUBBARD-HIGHTC}. 
It describes a lattice gas 
of itinerant electrons experiencing an on-site Coulomb interaction. 
So far, the evidence of pair-pair correlation is ambiguous
both for repulsive  and attractive Hubbard 
interaction\cite{MOREO-ZHANG,MARSIGLIO}.
Here we focus on  the Hubbard ($1d$) chains, with repulsive 
interaction\cite{HUBBARD-1D}. 
Not only does a $1d$ analysis
provide a unique theoretical lab to predict the behavior of 
compounds in higher dimension, but current technology 
actually admits quasi-$1d$ superconductors 
to be fabricated; finally, there is 
evidence that the stripe ordered domains gives effectively a $1d$ character to the phenomenology of even many  superconducting 
$3d$-compounds\cite{STRIPED}. 
However, how the superconductivity does arise from 
stripes still constitutes an open problem.    
Assuming that the  electronic properties 
of the stripe-domains  can be captured by repulsive 
Hubbard-like models, effective pair-formation can be  studied 
by means of the pair binding energy $E_{pb}$,
detecting an energy gain whenever an  electron-electron attraction 
is established into the system\cite{PAIR-BINDING,MARTINS,CHAKRA-KIVELSON}. 
In particular, $E_{pb}$ was studied for the ground state (GS) 
of the \textit{finite} $1d$ Hubbard  model and it turned out that  
the  repulsive local interaction can be over-screened, leaving an 
effective attraction among the electrons\cite{CHAKRA-KIVELSON}. 
This phenomenon was called mesoscopic pairing. 
An important motivation for us is to single out a  \textit{pairing
of the BCS type} in the  mesoscopic pairing of the Hubbard model. 
The attempt to trace BCS pairing within the Hubbard type models is not 
new, and it  has been  
well studied in the physical community, mainly for attractive 
interaction. A common feature of such studies\cite{MARSIGLIO} 
(with the exception of the variational study in Ref.
\onlinecite{MARSIGLIO-TANAKA})  is that the  analysis 
was  done in the grand-canonical ensemble, involving mean field BCS 
(that is exact in thermodynamic limit\cite{BOGOLIUBOV}). 
The scenario changes in the canonical ensemble where the BCS gap 
vanishes identically, the ``superconducting 
phase'' is dominated by quantum fluctuations\cite{MASTELLONE} and the mean 
field BCS theory  is inadequate\cite{OTHERS}. 
On the other hand, the canonical analysis 
is a natural tool to study  the mesoscopic BCS pairing in the
Hubbard model. This choice  in the context of the stripe physics is supported
by various experimental
evidences. Stripe ordered domains in superconducting nano-powders were detected
in Ref. \onlinecite{MOHANTY}; fragmented nano-stripes 
($10-20$ nanometers long) were clearly noticed also  in bulk high-T 
superconductors\cite{NANO-BULK-STRIPES}. 
We use  the Bethe ansatz solution of the BCS model to compare 
the two pictures in the canonical ensemble.
The exact solution of the BCS model was 
found by  Richardson~\cite{RICHARDSON} in 1963, 
but went unnoticed to the condensed matter community.
It was discovered only recently to   study 
the superconductivity at nanoscale\cite{MASTELLONE,OTHERS}.

The paper is organized in the following manner. 
In the next section 
we introduce the Hubbard model
and its Bethe Ansatz due to Lieb and Wu. In section \ref{sec:hubbard-gs}
a \textit{formal} correspondence is established for $U\to 0^+$,  
where the spin degrees of freedom (the spin ``quasi-momenta'') in the 
Hubbard GS can be seen as 
Cooper-pair quasi-energies corresponding to certain excitations of the BCS 
model. A finite pairing 
coupling can be achieved by considering certain  gauge-field 
correlated-hopping in the Hubbard model, as shown in the section 
\ref{sec:finite-pairing}. 
The correspondence of the BCS GS to a ``sea'' of 
spin singlets on top of the Hubbard GS, is discussed in section 
\ref{sec:bcs-gs}. 
The role of the BCS pairing in the mesoscopic pair-binding 
is discerned in the section \ref{sec:mesopairing}. Finally,
section \ref{sec:conclusions} is devoted to our conclusions, 
and the discussion of some physical implications.

\section{The Hubbard model}
\label{sec:hubbard-model}
We consider a Hubbard  chain of length $L$, with $N=N_{\up}+N_{\down}$ 
particles  $M=N_{\down}$ spin down, and periodic boundary conditions. 
Formulated in second quantization, the Hamiltonian is 
\begin{equation}
\label{eq:Hubbard}
H = - t \, \sum_{j=1 }^L 
(c_{j+1,\sigma}^\dagger c_{j,\sigma}+ h.c.) 
+ U\, \sum_{ j=1}^Ln_{{j},\uparrow} n_{{ j},\downarrow}\;,
\end{equation}
where  $\{c_{j,\sigma}, c_{l,\sigma'}^\dagger \}=\delta_{\sigma,\sigma'} \delta_{j,l}$,
$\{c_{j,\sigma}, c_{l,\sigma'} \}=0$, 
and $n_{l,\sigma}:= c_{l,\sigma}^\dagger c_{l,\sigma}$; $\sigma=\{\uparrow,\downarrow\}$  is the electronic spin; $U$ and 
$t$ (we set $t=1$) are the Coulomb interaction and the hopping amplitude 
respectively.
The many-body wave function 
\begin{equation}
\ket	{\psi} = \frac{1}{\sqrt{N!}} \sum_{1 \le j_1 \le \dots \le j_N \le L}
\sum_{\mathcal{S}_N} \psi (j_1, \dots, j_N | \pi) c^{\dagger}_{j_1 \sig_{\pi(1)}}
 \dots c^{\dagger}_{j_N \sig_{\pi(N)}} \ket{0},
\end{equation}
brings Eq. \eqref{eq:Hubbard} to the first quantized spectral 
problem ${\cal H} \psi={\cal E} \psi$, that can be solved by the Bethe 
ansatz\cite{FRAHM} 
\bea
\psi (j_1, \dots, j_N | \pi) &=& \sum_{Q \in \mathcal{S}_N} A_{\pi}(Q)
\exp \left \{i \sum_{l=1}^N k_{Q(l)} j_l \right \},
\\
A_{\pi}(Q) &=&
\sum_{R \in \mathcal{S}_M} G(R) \prod_{\alpha=1}^M F(\la_{R(\al)},y_\al|Q),
\\
F(\la,y|Q) &=& \frac{2iU}{\la - \sin k_y + iU} \prod_{l=1}^{y-1}
\frac{\la - \sin k_l - iU}{\la - \sin k_l + iU},
\\
G(R) &=& \prod_{1\le \al < \be \le M} \frac{\la_\al - \la_\be -2iU}
{\la_\al - \la_\be}.
\eea
In these equations $\pi$ labels a permutation of the indices in the
configuration space,
whereas $Q$ and $R$ label permutations in the $\{k\}$, $\{\la\}$
values in the symmetric group $\mathcal{S}_N$ and $\mathcal{S}_M$ respectively.

The exact energy and momentum of the  model\cite{HUBBARD-1D,FRAHM}
\begin{equation}
E=-2 \sum_{j=1}^N \cos k_j \; \qquad
P=\sum_{j=1}^N k_j 
\label{energy}
\end{equation}
are given in terms of  charge-momenta  $k_j$, solutions of the Lieb-Wu
 Bethe Equations (BE) \cite{LIEB-WU}
\begin{eqnarray}
e^{i L k_j}  &=&   \prod_{\alpha=1}^M 
\frac{ \sin k_j-\lambda_\alpha   + \frac{iU}{4}}
{ \sin k_j-\lambda_\alpha - \frac{iU}{4} } 
\label{eq:BAE1}
\\
\prod_{j=1}^N \frac{ \sin k_j-\lambda_\alpha + \frac{iU}{4}}
{ \sin k_j-\lambda_\alpha - \frac{iU}{4} }  &=& 
\prod_{\substack{\be=1 \\ \be \neq \al}}^M \frac { \lambda_{\alpha} - \lambda_{\beta}- 
\frac{iU}{2} }{ \lambda_{\alpha} - \lambda_{\beta} + \frac{iU}{2} } ,
\label{eq:BAE2}
\end{eqnarray}
where  $ j \in \{ 1,2, \dots N\}$ and $\alpha \in \{1,2, \dots M \}$.

The equations \eqref{energy}, \eqref{eq:BAE1}, and \eqref{eq:BAE2}
constitute the exact solution of the 
Hubbard model both for repulsive and attractive interactions.  
We focus on the limit $U\to 0^+$.
The result of such operation on Eqs. 
\eqref{energy}, \eqref{eq:BAE1} and \eqref{eq:BAE2} 
depends on the actual configurations of the charge 
and spin rapidities $k,\lambda$. For numerics,
systems up to $N=100$ electrons have been considered.

\section{Hubbard ground state for vanishing interaction}
\label{sec:hubbard-gs}

\begin{figure}
\includegraphics[width=7cm]{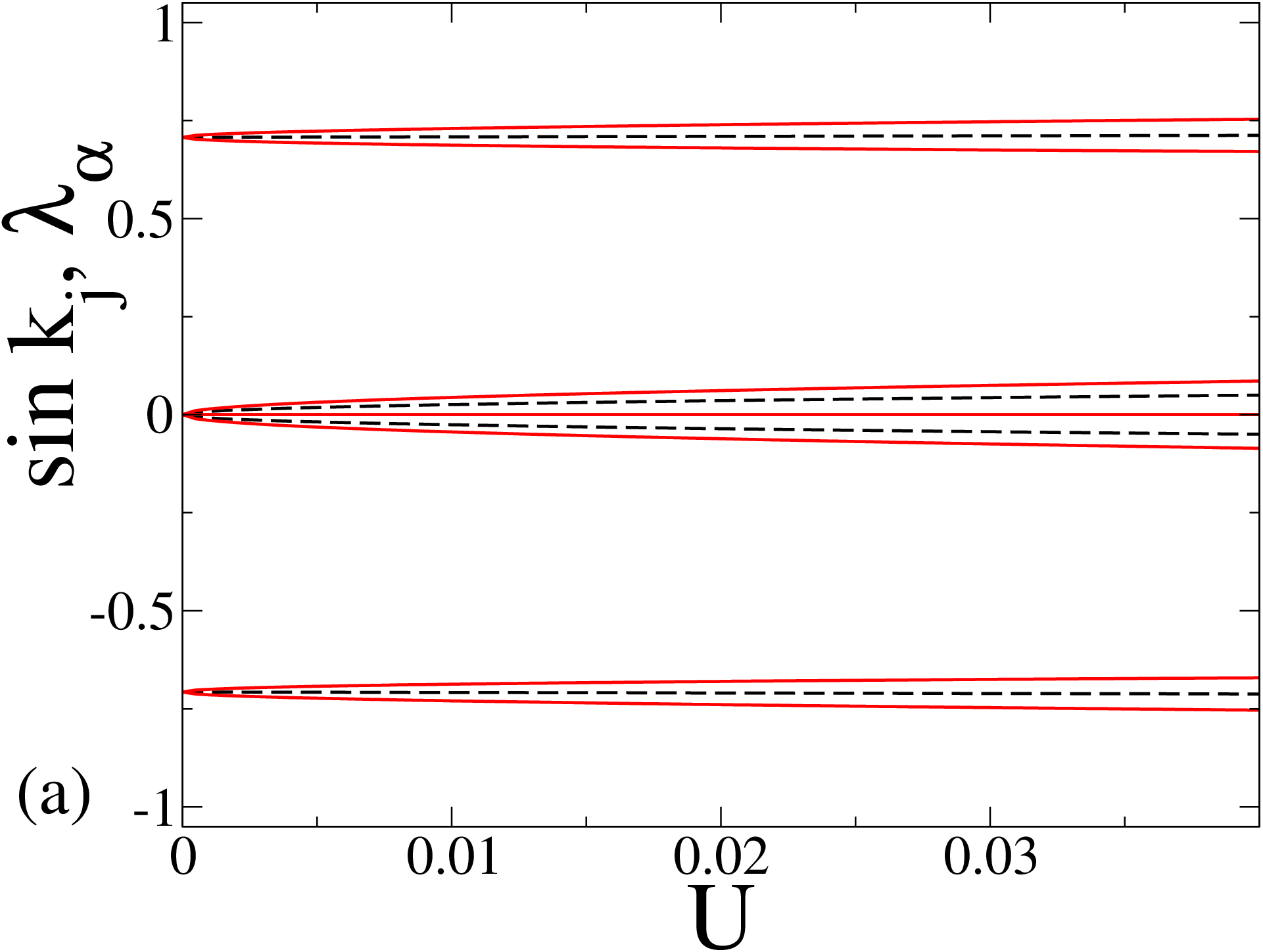}
\includegraphics[width=7cm]{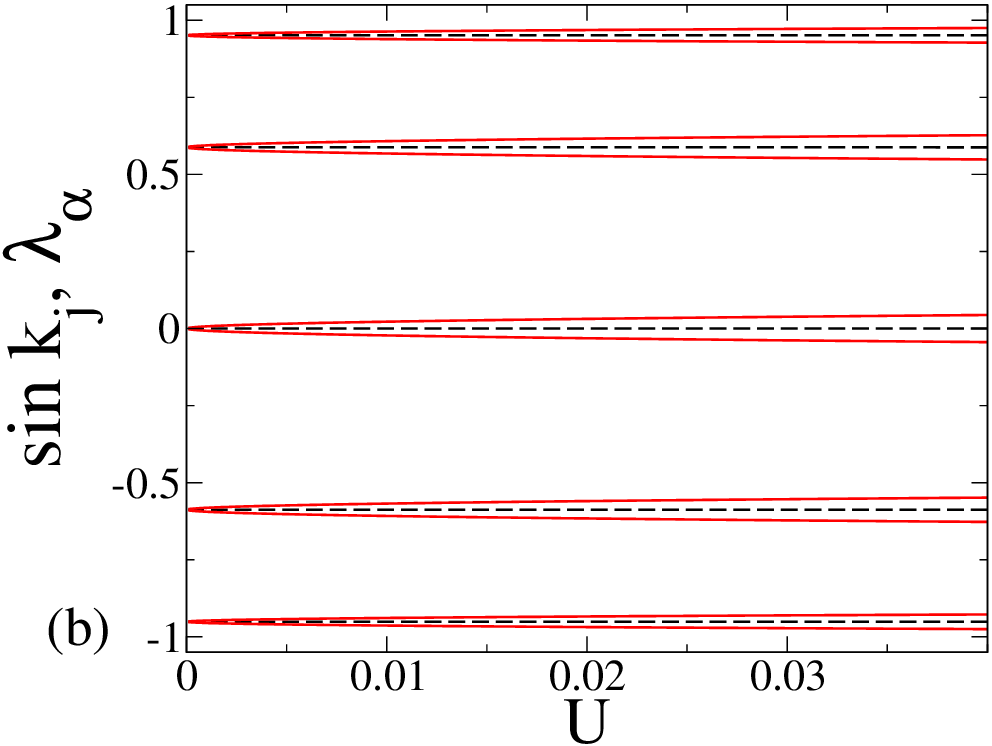}
\includegraphics[width=7cm]{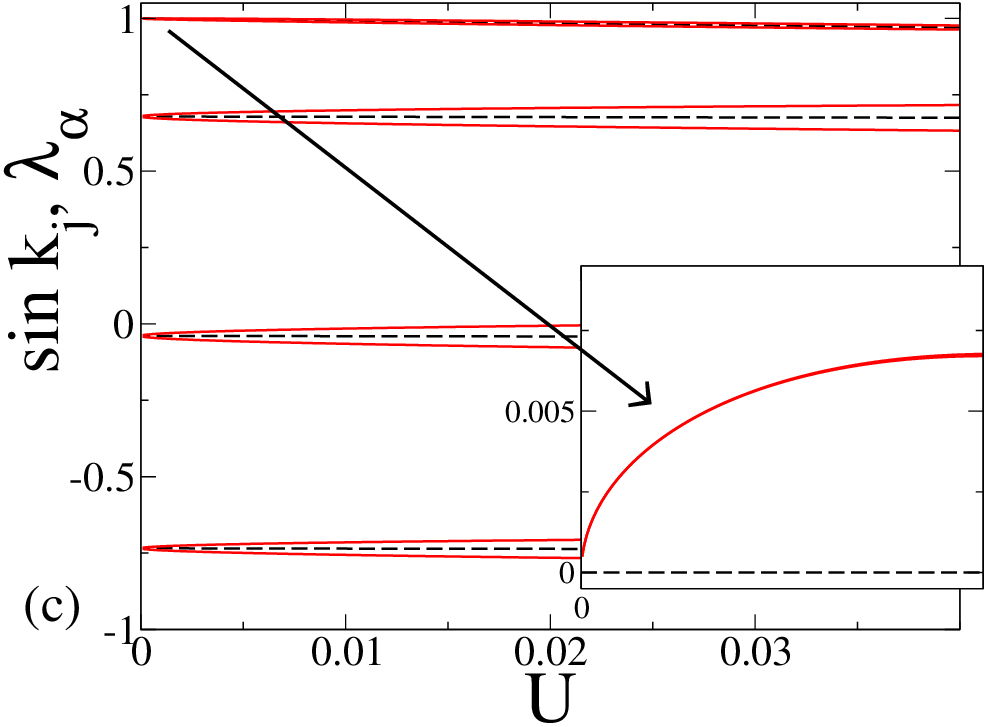}
\caption{The spin ($\la_{\al}$, black dashed lines) 
and charge ($\sin k_j$, red solid lines) rapidities
at  $U/t \ll 1$, for a system with 8 (a) and 10 (b) electrons. 
A correlated hopping $\Phi_\sigma \neq 0$ removes the degeneration 
of the cluster at $k=0$ for $N=4n$ (c). 
In the upper cluster, though $\la_\al$ exhibits a slight dependence 
on $U$, the functional behavior of the difference $|\la_\al - \sin k_\al|$
remains the same (inset).}
\label{fig:ground}
\end{figure}

Here we consider the half-filled chain, and   $N = 2M$, 
$N$ even. For $U>0$ the GS is characterized 
by real $k$ and $\lambda $, where any $\lambda_\al$ is trapped between two 
$\sin k_\alpha$.  
Systems with $N=4n$ electrons  are characterized by degenerate 
$\sin {k}=0$ (see the central cluster of Fig. \ref{fig:ground} (a)).
At small $U$ the two $\sin k_\al$'s are arranged symmetrically around 
$\lambda_\al$ as $|\lambda_j-\sin k_j| \simeq C_j U^{1/2} + D_j U$ 
where $C_j$ and $D_j$ are numerical coefficients; 
from the numerical analysis we find $D_j=1/4$.
We further assume the quasi-momenta having the following behavior
$k_j \simeq k_j^{(0)} + U^{1/2} k_j^{(1/2)} + U k_j^{(1)}$; 
this assumption will be confirmed by our numerical studies. 
The limit $U \rightarrow 0^+$ of Eqs. \eqref{eq:BAE1} and \eqref{eq:BAE2}
reads~\cite{FRAHM,BATCHELOR}
\bea
&& e^{i k_j^{(0)} L} = 1\; , 
\label{eq:group1order0} 
\\
&&k_j^{(1)} L = \frac{1}{2} \sum_{\alpha} \, '  
\frac{1}{\sin k_j^{(0)} -\la_{\al}},
\label{eq:group1order1im} \\
&&\sum_{\be \neq \al} \frac{2}{\lambda_{\beta} - \lambda_{\alpha}}
- \sum_{j=1}^N  \frac{1}{\sin k_j - \la_{\al}}  =  0.
\label{eq:group2order1}
\eea
and $k_j^{(1/2)} L = {1}/{2 C_j}$. 
$\sum'$ in Eq. \eqref{eq:group1order1im} indicates that the term 
$\sin k_{\al}$ is omitted. 
Notice that $U$ has to be small in order 
that both $|U^{1/2} k_j^{(1/2)} L| \ll 1$ and $|U k_j^{(1)} L| \ll 1$
are satisfied and  that degenerate $\sin k_j$ may occur in 
Eq. \eqref{eq:group2order1} (see Fig. \ref{fig:ground}).

The Bethe ansatz wavefunction in the limit of vanishing $U$ is\cite{FRAHM}
\beq
\lim_{U\rightarrow 0} \frac{A_{\pi}(Q)}{(2iU)^M} = 
\sum_{R\in \mathcal{S}_M} \prod_{\al=1}^M \frac{1}{\la_{R(\al)}-\sin[k_{Q(y_\al)}]}
\label{eq:wavefun-zerolimit}
\eeq

The relevance of this limit resides in the following observation.  
The Eqs. \eqref{eq:group2order1} are  known as the Gaudin equations.
They play an important role in the exact solution of the BCS 
model at infinite pairing coupling $g$. 
The  BCS Hamiltonian  for $N_p$ time-reversed pairs 
in $\Omega$ single particle levels $\varepsilon_j$ with pair degeneracy $d_j$ reads
\cite{SIERRA-RMP}
\bea
\label{BCS-ham}
&&\hspace*{-0.4cm}H_{_{BCS}} =  
\sum_m \sum_{j=1}^\Omega \varepsilon_j c^\dagger_{j,m} c_{j,m}-g
\sum_{i,j=1}^\Omega \sum_{m,m'}  c^\dagger_{i\,m} c^\dagger_{i\,\bar{m}}
c_{j\,\bar{m'}}c_{j\,m'} \nonumber \\ &&\hspace*{0.5cm}  = 
\sum_{j=1}^\Omega \varepsilon_j \tau_j \;, \quad
\tau_j = \sum_{i \neq j} \frac{\vec{S}_i \cdot \vec{S}_j}
{\varepsilon_i - \varepsilon_j} + \frac{1}{g} S^z_j,
\eea
where $c^\dagger_{j\,m}$ creates an electron in the state 
$\ket{j;S_j,m}$, $m\in\{1\dots d_j\}$; 
$|j;S_j,\bar{m}\rangle$ is  the time reversed state of $|j;S_j,m\rangle$. 
The operators 
$
S^-_j := \sum_m c_{j\overline{m}} c_{jm}, \;\; S^+_j:=(S^-_j)^\dagger \;\;
S^z_j :=  (\sum_{m=-d_j}^{d_j} n_{jm} -d_j)/2$ form a $d_j/2$ dimensional
representation of the $su(2)$ algebra. 
The second line of Eq. \eqref{BCS-ham} establishes the relation 
between $H_{BCS}$ and the twisted six-vertex models\cite{AMICO-VERTEX}  
through the Gaudin model\cite{CAMBIAGGIO},  ultimately 
proving the integrability of the BCS model. 
Its  exact energy and eigenstates\cite{RICHARDSON}  are 
\begin{eqnarray}
&&\mathcal{E}_{BCS}  = \sum_{\alpha =   1}^{N_p} e_\alpha 
+ \sum_{i=1}^{N_b} \varepsilon_i \label{BCS-exact-en}\\ 
&&\frac{1}{g} - \sum_{j=1}^\Omega \frac{d_j}{2 \varepsilon_j -e_\al}+ 
\sum_{\be\not=\al=1}^{N_p} \frac{2}{e_\be -e_\al}=0 \;,
\label{BAE_BCS}\\
&&|\Psi\rangle =\sum_{R\in \mathcal{S}_{N_p}} \prod_{\al=1}^{N_p} \frac{1}{e_{R(\al)}-
2\varepsilon_{Q(y_\al)}}S^\dagger_{R(\al)}=\prod_{\al=1}^{N_p} \sum_{j=1}^\Omega 
\frac{S^\dagger_j}{ e_\al-2 \varepsilon_j } \ket{0} \label{BCS-exact-state}\end{eqnarray}
where 
the second sum in Eq.\eqref{BCS-exact-en} runs over the set of $N_b$
blocked levels, i.e. occupied by a single electron
and not available to pair scattering.
Different states in Eq. \eqref{BCS-exact-state} correspond to different 
distributions of the $e_\al$ at $g=0$ on the $\varepsilon_j$-configuration; 
in the BCS GS all $\varepsilon_j$'s are complex conjugates.
The relation between  Eqs. \eqref{BAE_BCS},\eqref{BCS-exact-state}  
and Eqs. \eqref{eq:group2order1}, \eqref{eq:wavefun-zerolimit} is evident
identifying $2 \varepsilon_j$ and $e_\al$ 
(the pairing quasi-energies) with   $\sin k_j$ and $\lambda_\al$ 
respectively, 
$\Omega$ and $N_p$ playing the role of $N$ and $M$. 
Eqs. \eqref{eq:group2order1} are obtained from Eq. \eqref{BAE_BCS} 
in the limit $g\rightarrow \infty$. 
Important for our purposes are the excitations   
of the BCS model characterized by real $\{ e_\alpha\}$; it was noticed
that in the canonical ensemble these can be obtained as the 
states in Eq. \eqref{BCS-exact-state}
corresponding to $\{ e_\alpha\}$, \textit{finite} solutions of the Gaudin 
equations\cite{SIERRA-RMP}. Therefore we conclude that the 
spin-rapidities (all of them are real and finite) in the Hubbard GS are 
arranged along the configuration of the BCS  excited state
 displayed in the Fig. \ref{fig:bcs-conf}.

\section{Finite pairing}
\label{sec:finite-pairing}

Now we show that a finite value of $g$ can be obtained by adding  
a further interaction in the 
original Hubbard model, in  form of correlated-hopping. 
The correlated hopping we consider corresponds to 
\begin{equation}
t \rightarrow t \exp\Bigl[{\rm i}\sum_{l}^{}\bigl(\alpha_{j,l}(\sigma)N_{l,-\sigma}
   +  A_{j,l}(\sigma)N_{l,\sigma}\bigr)\Bigr] \; ,
\label{correlated}
\end{equation}
in Eq. \eqref{eq:Hubbard}. 
The Hubbard model with the correlated  hopping Eq. \eqref{correlated} are 
of the Shastry-Schulz type\cite{SCHULZ-SHASTRY}. 
They were solved exactly for  $A, \alpha$ obeying certain 
restrictions\cite{AMICO-OSTERLOH-ECKERN}. 
It was demonstrated \cite{AMICO-OSTERLOH-ECKERN} that 
the effect of the correlated hopping is to twist the boundary conditions.  
The boundary phases are 
$\Phi_\sigma:=\phi(\sigma)+ \theta_{\uparrow\downarrow}(\sigma) 
N_{-\sigma} +\theta_{\uparrow\uparrow}(\sigma)(N_\sigma-1)$,
with $ \phi(\sigma)
=\sum_{j=1}^{L} A_{j,j}(\sigma)$,
$\theta_{\uparrow\downarrow}(\sigma)
= \sum_{j=1}^{L}\alpha_{j,m}(\sigma)$, 
$ \theta_{\uparrow\uparrow}(\sigma)
=\sum_{{{j\neq m-1,m}}}^{L} A_{j,m}(\sigma)
+A_{m,m-1}(\sigma)+A_{m-1,m+1}(\sigma)$.
The BE of the Schulz-Shastry models are obtained from the Lieb-Wu BE, 
by  multiplying with 
the factors $e^{-i\Phi_{\uparrow}}$ and $e^{i(\Phi_{\uparrow}-\Phi_{\downarrow})}$ 
the r.h.s. of the equations \eqref{eq:BAE1} and \eqref{eq:BAE2} respectively.
To obtain a finite $g$, it is assumed  also that 
$\Phi_{\sigma} \simeq \Phi_{\sigma}^{(0)} + U \Phi_{\sigma}^{(1)}$.
For $-\pi \le k_j < \pi$ we find that the conditions $\Phi_{\up \down}^{(0)}=
\Phi_{\up}^{(0)} - \Phi_{\down}^{(0)} = 0$ and $|U \Phi_{\up \down}^{(1)}| \ll 1$
must be satisfied, in order to obtain the BE to zero order in $U$. 
The Eqs. \eqref{eq:group1order0} and \eqref{eq:group1order1im} then modify 
into 
\bea
e^{i k_j^{(0)} L} &=& e^{-i \Phi_{\up}^{(0)}},
\nonumber
\label{eq:newgroup1order0}
\\
k_j^{(1)} L &=& - \Phi_{\up}^{(1)}
+ \frac{1}{2} \sum_{\alpha} \, '  \frac{1}{\sin k_j^{(0)} -\lambda_\alpha},
\nonumber
\label{eq:newgroup1order1im}
\eea
and  Eqs.(\ref{eq:group2order1}) into 
\begin{equation}
\frac{1}{g} - \sum_{j=1}^N \frac{1}{\sin k_j -\lambda_\al}+ 
\sum_{\be\not=\al=1}^{M} \frac{2}{\lambda_\be -\lambda_\al}=0 \;,
\label{eq:pairing}
\end{equation}
with 
$ g \equiv 1/(2\Phi_{\up \down}^{(1)})$.
Thus the BE of the correlated  hopping Hubbard model lead, 
in the limit $U \rightarrow 0^{+}$, to Richardson BE, Eq. \eqref{BAE_BCS}.
In the same limit the Eq. \eqref{energy} is 
$$
E = \frac{2 U \Phi_{\up \down}^{(1)}}{L} \sum_{\al=1}^M \la_{\al}
+ \mathrm{const}.
$$
with the constant being ${U M(N-M+1)}/{L} -2  \sum_{j=1}^N \cos k_j^{(0)}
 - ({2 U \Phi_{\up}^{(1)}}/{L}) \sum_{j=1}^N \sin k_j^{(0)}$
That is: for small $U$, the energy of the Hubbard model  coincides (up to constants) 
with the energy of the BCS model 
, see Eq.\eqref{BCS-exact-en}.

\begin{figure}
\includegraphics[width=7cm]{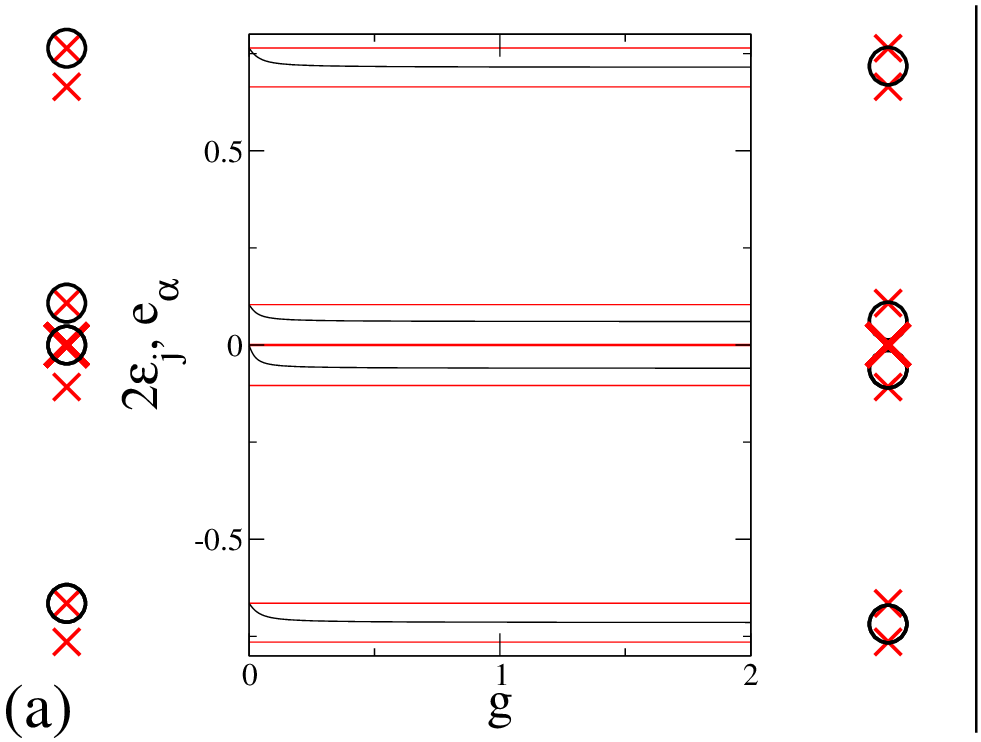}
\quad
\includegraphics[width=7cm]{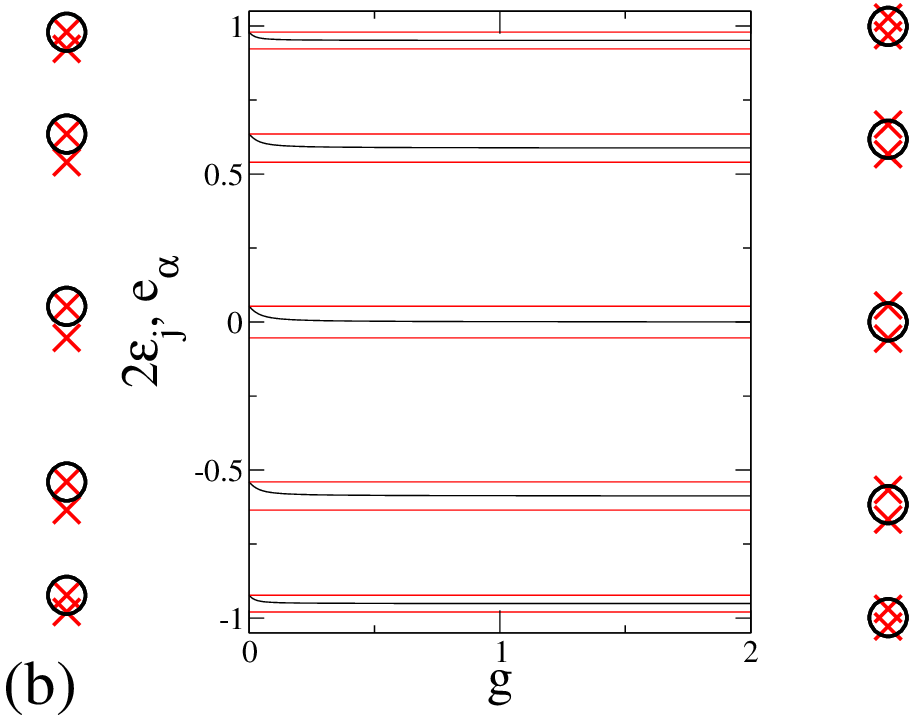}
\caption{BCS and Gaudin configurations leading to finite solutions
at large pairing, for $N=4n$ (a) and $N=4n+2$ (b) and $n=2$.
The plot
shows the behavior of the pairing parameters while increasing $g$,
black curves, whereas the single particle energies remain constant,
red lines.
On left sides of the plots, the initial configuration of the
pairing parameters $\la_\al$ (black circles) on the
energy levels $2\varepsilon_j$ (red crosses) is shown;  
on the right sides the final configuration is displayed. 
The latter are the charge and spin configurations corresponding to 
the ground state of the Hubbard model. The larger cross in plot (a) 
indicates the double degenerate level $2\varepsilon=0$.} 
\label{fig:bcs-conf}
\end{figure}

The excitation of  the BCS model corresponding to the  Hubbard GS  
is achieved by filling the levels  
$\varepsilon_j$'s with a Cooper pair, leaving one empty 
level between two filled ones. 
In this way, the empty
levels prevent the pairing parameters to form complex 
conjugate pairs. Such a BCS  excitation is constituted by  an 
 non-condensed liquid of Cooper pairs (see the discussion in the next 
paragraph).  
The structure of these excitations is different for $N=4n+2$ and $N=4n$. 
In the first case the $\varepsilon_j$ are all non degenerate: 
$d_j=1,\; \forall j $; for $N=4n$, instead, the doubly degenerate Fermi  
energy correspond to a doubly degenerate $\varepsilon_j$, as one can see in
the Figs. \ref{fig:ground} and \ref{fig:bcs-conf}).

\section{The BCS ground state}
\label{sec:bcs-gs}

\begin{figure}
\includegraphics[width=10cm]{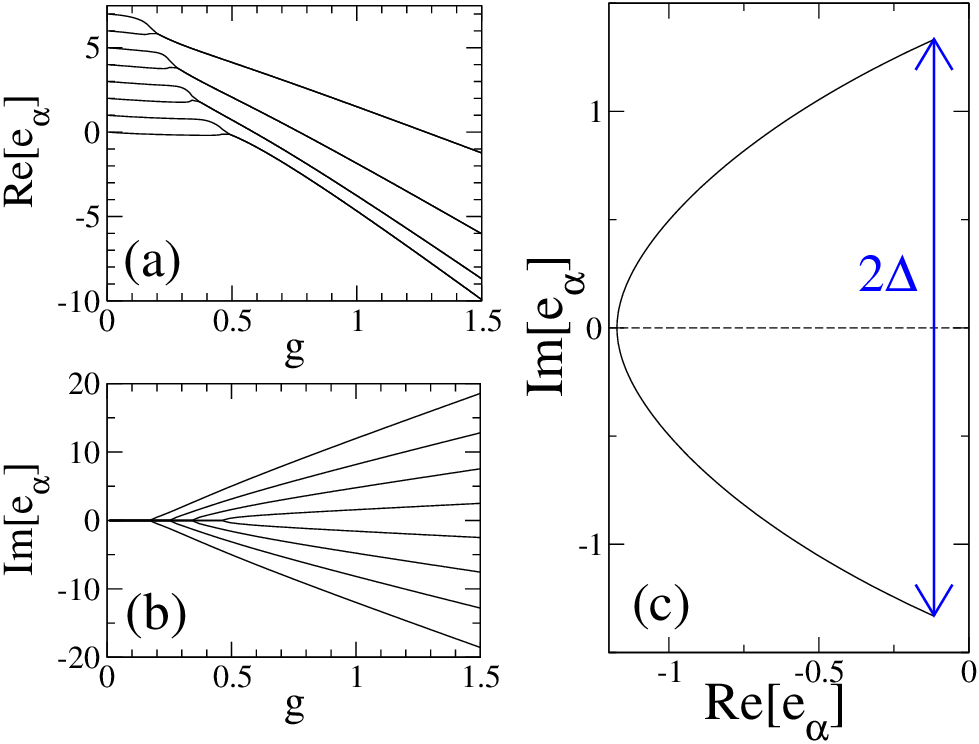}
\caption{(a) The real and (b) the imaginary part of the pairing 
parameters are shown as functions of the pairing strength $g$
for a system with $N=16$ electrons at half-filling in the BCS
ground state configuration. In the thermodynamic limit the pairing
parameters, being all complex conjugate at sufficiently large pairing,
are arranged along arcs (see Ref. \onlinecite{ELECTRO}), 
where the distance between the end-points
is the BCS gap $2\Delta$. Here we have set $g=3.0$ and $N=200$, with 
the single particle levels uniformly distributed in the range [-1,1].}
\label{fig:pair-parameters-conf}
\end{figure}

The trait of the condensation of the Cooper pairs in the  
BCS GS is the emergence of 
complex conjugate $e_\alpha$, solution of Eq. \eqref{BAE_BCS}. 
In fact by resorting  certain electrostatic analogy 
\cite{GAUDIN,RICHARDSON-ELECTRO} it can be proved that 
the thermodynamic limit of the Richardson's BE 
leads to the gap equation: $2G \int_\Omega \rho(\varepsilon)/
\sqrt{(\varepsilon-a) (\varepsilon-b)}=1$ where $\rho(\varepsilon)$ is the density of single particle levels; the parameters $a=\varepsilon_0+i\Delta$ and 
$b=\varepsilon_0-i\Delta$ are the the end points of the arc where the solutions of Eqs.(\ref{BAE_BCS}) are disposed (see  
Fig.\ref{fig:pair-parameters-conf}; see also Refs.
\onlinecite{SIERRA-RMP,ELECTRO}). 
Thus: the BCS gap (in the thermodynamic limit) 
is directly related to the maximum  of the imaginary part of the
$e_\alpha$\cite{RICHARDSON-ELECTRO,SIERRA-RMP,ELECTRO}. 
For large $g$ all the $e_\alpha$ are complex 
conjugated pairs (with a single exception if $N_p$ is odd); 
their real part is far below the lowest $\varepsilon_j$
(see e.g. Ref. \onlinecite{SIERRA-RMP}). 

To  find the Hubbard eigenstate corresponding 
to the BCS GS we insert into Eqs. \eqref{eq:BAE1}, \eqref{eq:BAE2} 
the $e_\alpha$($\equiv \lambda_\alpha$) configuration
obtained by solving Eq. \eqref{eq:pairing} with certain configuration of  
$ \sin k_j^{(0)}$; this fixes the $k_j$, given the $\Phi_\sigma$
we have used for Eq. \eqref{eq:pairing} 
with  $|U \Phi_{\sigma}^{(1)}| \ll 1$. 
We found that all the $k_j$'s 
are real and non degenerate for non degenerate $\sin k_j^{(0)}$'s.
(degenerate $\sin k_j^{(0)}$ correspond to degenerate $k_j$).
Numerically  we found  that  the mapping can be reliably done for  
$U \lesssim 0.03$. It then turns out that 
$\sin k_j=\sin k_j^{(0)} + U \delta_j$; 
and $|\sin k-\lambda|\neq 0$ is always satisfied. These conditions 
imply that equation (\ref{eq:pairing}) characterizes also the 
excited Hubbard state under consideration.
Summarizing, real $k_j$ $\&$ complex conjugate $\la_{\al}$ pairs, 
solutions of the Richardson BE, satisfy  the Lieb-Wu equations. 
We conclude that the BCS GS corresponds to a sea of 
spin singlets ($\Lambda$-2 strings) in the Hubbard chain. 
Such a state can be seen  as a BCS
condensation of spin rapidities. 
As an example of such result, the table shows the $k,\lambda$ solution 
of both the BCS and the Hubbard GS BE for $U=0.001$, $N=6$, and $M=3$. 

\begin{center}
\begin{tabular}{|cc|c|}
\hline
$\sin k_j$  &  & $\lambda_\alpha $ \\
\hline
\small
 -2.14679391, & \small -1.09959633  &  \small -14.8060834 \\ 
\hline
\small -0.052399128, &  \small 0.994798057  &  \small -11.722088 - i 11.1576888 \\ 
\hline
\small 2.04199559,   & \small  3.08919348 &  \small -11.722088 + i 11.1576888 \\ 
\hline 
\end{tabular}
\normalsize
\end{center}

\section{Mesoscopic pairing}
\label{sec:mesopairing}

\begin{figure}
\includegraphics[width=8cm]{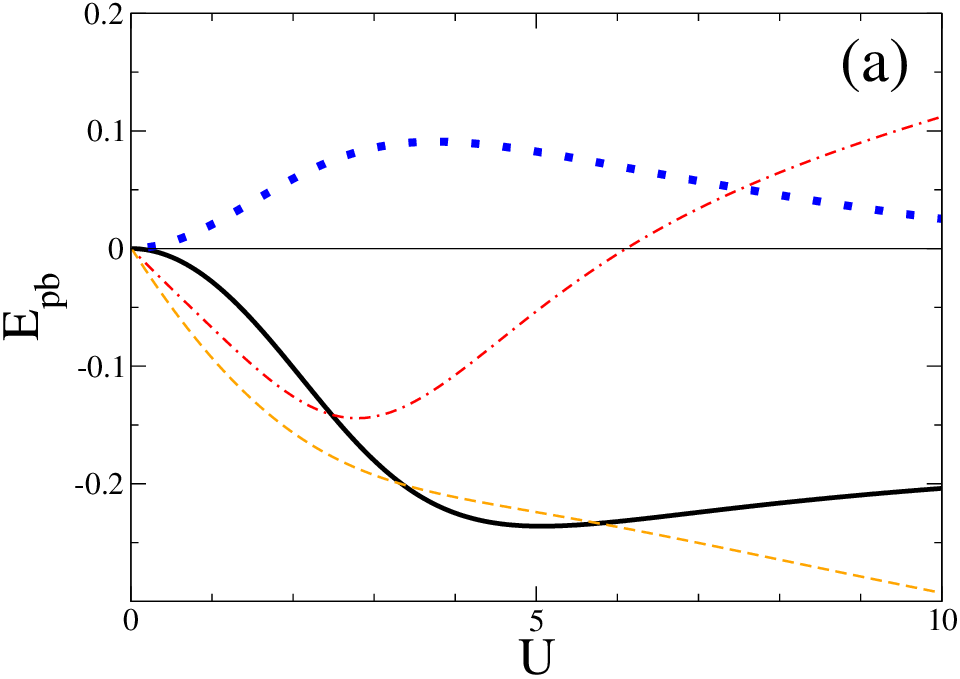}
\includegraphics[width=8cm]{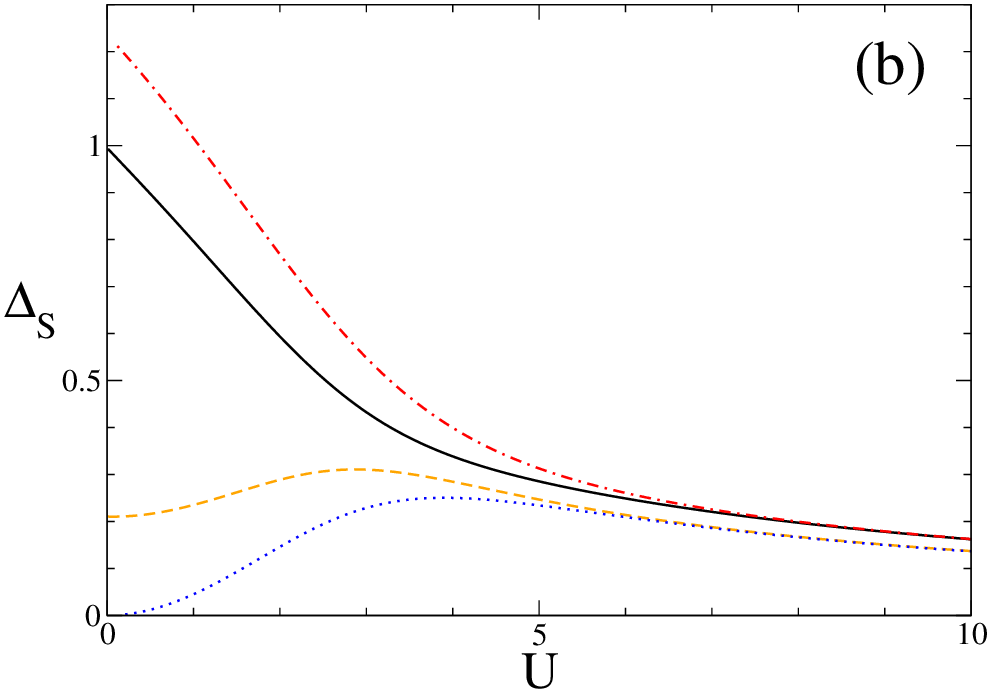}
\caption{The super-even effects for the pair binding energy $E_pb$ (a) and 
the spin gap $\Delta_s$ (b).
Thick and light lines are results for periodic and twisted boundaries 
respectively.    
$N=10$: black solid line and red dotted-dashed line; 
$N=12$: blue dotted line and orange dashed line.
The boundary twist is set to  $\Phi^{(0)}_{\up} = \Phi^{(0)}_{\down} = 0.2 \pi$,
$\Phi^{(1)}_{\up} = 0.1 \pi=2\Phi^{(1)}_{\down}$. }
\label{fig:bindspin}
\end{figure}

\begin{figure}
\includegraphics[width=4cm]{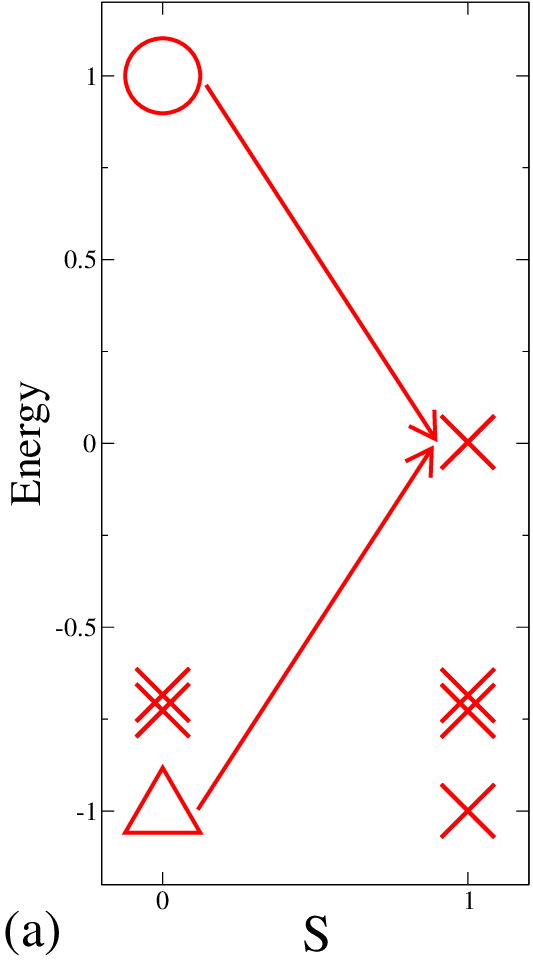}
\quad
\includegraphics[width=4cm]{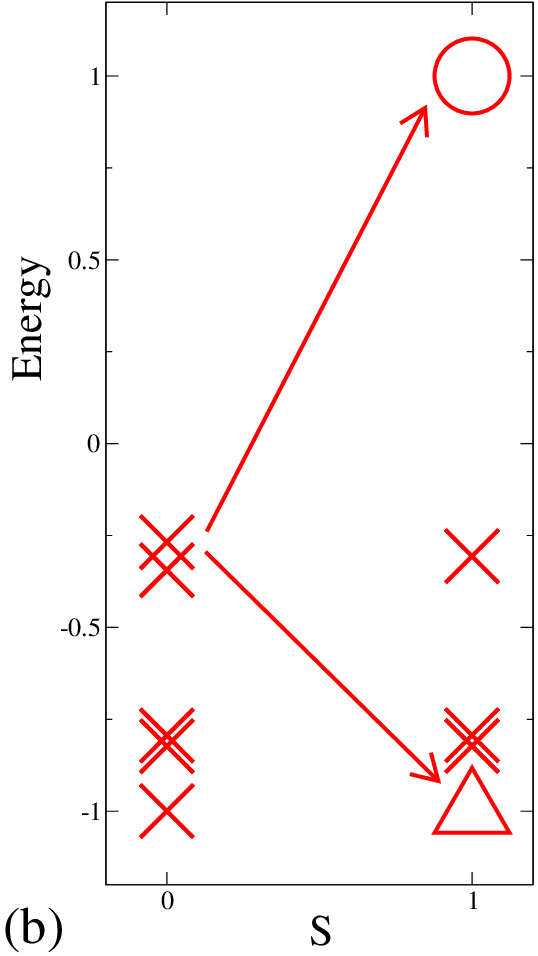}
\caption{The energy configuration ($-2\cos k$) 
of the charge rapidities in the $S=0,1$ ground states for $N=4n$ 
(a) and $N=4n+2$ (b) systems. 
The different symbols indicate the degeneracy of 
the levels: circle, cross and triangle stand for one, two and 
three-fold degenerate levels respectively. Note that 
the three-fold degenerate levels are more precisely ``quasi-degenerate''
ones: the energy difference cannot be resolved in the plot. 
At $N=4n$, passing from $S=0$ to $S=1$ state two degenerate single
particle levels are created and no additional energy is required (as
shown by the arrows). Instead, at $N=4n+2$, at half filling the
levels are reshuffled, a pair is broken at Fermi level and a $k=0$ 
level is occupied (upper arrow).} 
\label{fig:energy-conf}
\end{figure}

Here, we have a look at the pair binding energy 
$E_{pb} \doteq 2 E(N+1) -E(N) -E(N+2)$. 
For the Hubbard  GS,   
$E_{pb}$ shows ``super-even effects''\cite{CHAKRA-KIVELSON,PAIR-BINDING}. 
Effective pair attraction $E_{pb}>0$ occurs  for $N=4n$ ($n$ is integer), and it is 
related with a  vanishing GS spin gap $\Delta_S\doteq E(N)_{S=1}-E(N)_{S=0}$ 
for $U \rightarrow 0^+$; 
for $N=4n +2$, $E_{pb}<0$, corresponding to $\Delta_S>0$
at  small $U$. In Fig.(\ref{fig:bindspin}) we present $E_{pb}$
for the Hubbard and for Shastry-Schulz models. We observe that    
the correlated hopping, Eq. \eqref{correlated}, washes out
the super-even effects 
(see Fig.\ref{fig:bindspin}); for $N=4n+2$ and $U> U_c$ $E_p>0$
(see also Ref. \onlinecite{MARTINS});
at small $N$ we found that $U_c \propto N$. For larger $N$, 
$U_c$ decreases (at $N=50$, $U_c\approx 2.5$). The phenomenon will be analyzed elsewhere. 
The  formation of $\Delta_S$ can be interpreted  
within the Hubbard-BCS  correspondence:
the state with $S=1$ is 
obtained from $S=0$ by breaking a BCS pair and thereby blocking 
single particle levels. In this scenario the super-even effect arises
(see also Fig.\ref{fig:energy-conf}) because breaking a pair 
in the state with  $N=4n$  is energetically favorable, creating two 
blocked degenerate single particle levels, without kinetic energy extra cost. 
In contrast, for $N=4n+2$ the blocking of  levels occurs at $k>k_F$.
 Away from half filling this is sufficient to cause  an increase of the total energy; 
at half filling the levels are blocked 
in such a way that all the  levels in the band are reshuffled, 
since breaking a pair at the Fermi energy causes the occupancy of the state at $k=0$.
We study the phenomenon by considering the pair 
correlation function $\Psi_j=|u_j|$, where 
$u_j\doteq \langle c^\dagger_{j\,m} c^\dagger_{j\,\bar{m}} c_{j\,\bar{m}}c_{j\,m}\rangle $ 
detecting the fluctuational superconductivity in the canonical ensemble  
(in the thermodynamic limit $\Psi_j$ becomes the BCS gap)\cite{MASTELLONE}. 
Here we exploit the exact results achieved in 
Ref. \onlinecite{BCS-CORR} where  $\Psi_j=|u_j|$ is calculated 
from a generating function\cite{SKLYANIN-CORR}. 

\begin{figure}
\includegraphics[width=10cm]{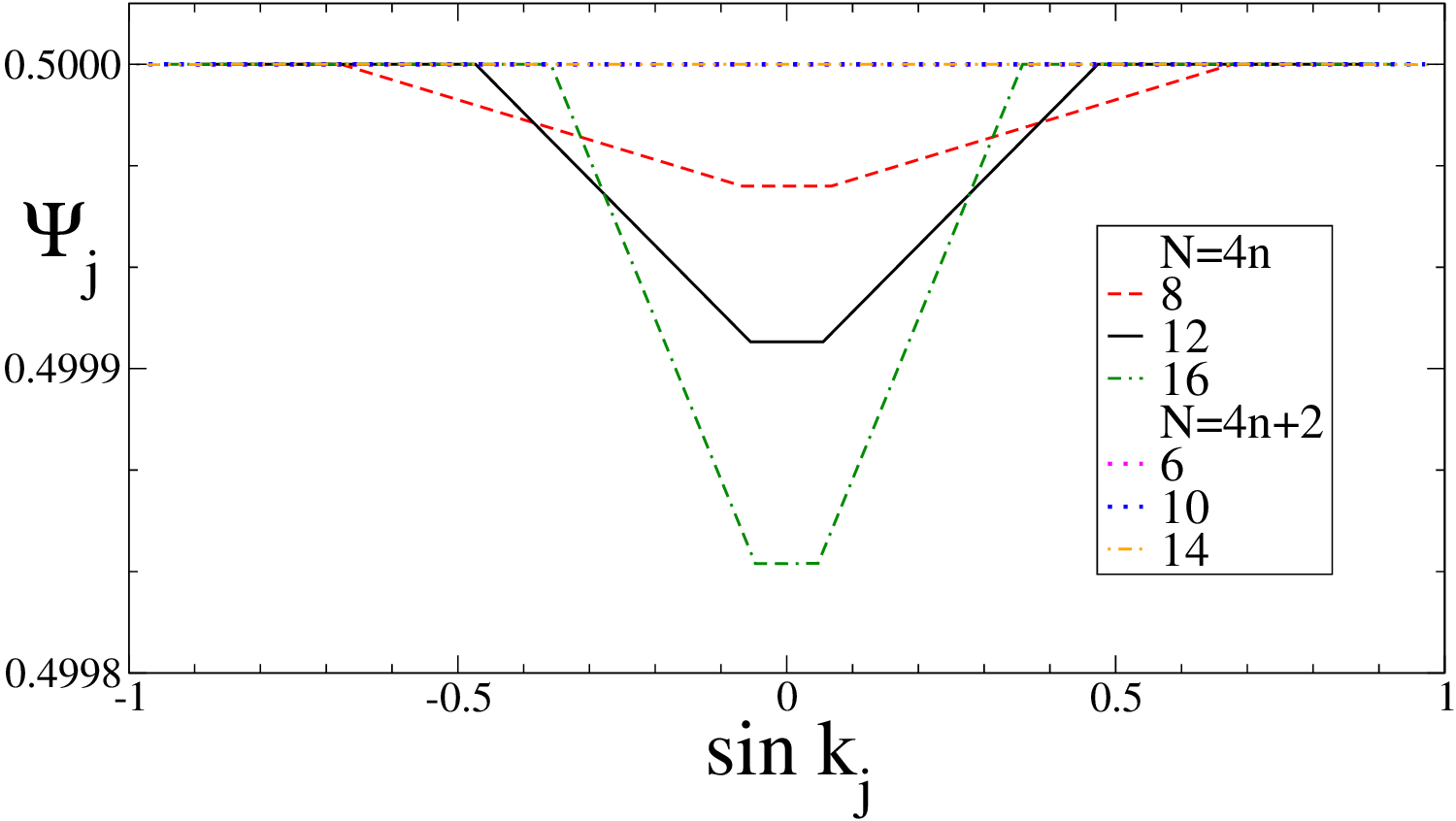}
\caption{The pairing correlator  $\Psi_j=|u_j|$
is presented (in the thermodynamic limit $\Psi_j$ would become the BCS gap). 
We notice the anomaly for $N=4n$.}
\label{fig:paircorr}
\end{figure}

Exclusively for 
systems with pair attraction ($N=4n$ with $\Phi_\sigma\equiv 0$)  
a dip at $k=0$ is observed (Fig.\ref{fig:paircorr}); this reflects the
corresponding small energy gap required to occupy the $k=0$ mode in 
the  $S=0\rightarrow S=1$ process. 

\section{Conclusions}
\label{sec:conclusions}

We proved that the exact solution of the BCS model 
can be obtained from the Bethe ansatz solution of the Hubbard model at 
vanishing $U$. At a formal level we comment that  the amplitudes $A_\pi(Q)$ diagonalizing the 'spin sector'  of the Hubbard model are, in fact, eigenvectors of the transfer matrix of certain inhomogeneous six vertex models\cite{FRAHM}.
The relation between the Hubbard chain and the BCS model arises 
because the  BCS model is itself a quasi-classical descendent of 
the inhomogeneous six vertex model\cite{AMICO-VERTEX}. 
The Coulomb repulsion $U/t$ plays the role of the quasi-classical parameter; 
the charge degrees of freedom $\{\sin k \}$ of the Hubbard model play the role 
of the inhomogeneities for the vertex models.  In the 
BCS picture, they provide  
the ``lattice'' of single particle energies  $\varepsilon_j$ where the spin 
degrees of freedom, the spin ``quasi-momenta''  $\lambda_\alpha$, 
can condense.
 The Hubbard ground state corresponds to a certain  
excited BCS state (see  Fig.\ref{fig:bcs-conf}).
We noticed that also the Hubbard low lying spin excitations corresponds 
to states of the BCS type. 
In particular we have demonstrated that the BCS ground state of the spin rapidities 
reflects the formation of a sea of spin singlets in the Hubbard chain. 
In this scenario  
the super-even effective attraction in the Hubbard  and in the Shastry-Schulz
ground states can be understood in terms of pair-breaking excitations 
in the BCS model.  We have demonstrated that BCS-correlators  serve to study  
such parity effects.
 
The  results obtained in this paper might be relevant for the analysis of 
the stripe order in the high-$T_c$ compounds.

\begin{acknowledgments}
A. Di Lorenzo, F. Dolcini, G. Falci, R. Fazio, H. Frahm, A. Fubini, 
M. Roncaglia, and G. Sierra are 
acknowledged  for discussions and support. 
\end{acknowledgments}

\end{document}